%% file: main.tex
\documentclass[10pt, journal, draftclsnofoot, onecolumn]{IEEEtran} 
\usepackage[T1]{fontenc}
\IEEEoverridecommandlockouts
\usepackage{authblk}
\usepackage{cite}
\usepackage{amsmath,amssymb,amsfonts,dsfont}
\usepackage{algorithm}
\usepackage[noend]{algorithmic}
\usepackage{graphicx}
\usepackage{textcomp}
\usepackage{xcolor}
\usepackage{bm}
\usepackage{booktabs}
\usepackage{multirow}
\usepackage{lipsum}
\usepackage{url}

\ifCLASSOPTIONcompsoc
\usepackage[caption=false, font=normalsize, labelfont=sf, textfont=sf]{subfig}
\else
\usepackage[caption=false, font=footnotesize]{subfig}

\newtheorem{theorems}{\textbf{Theorem}}

\newtheorem{lemma}{\textbf{Lemma}}

\input{notation}

\newenvironment{proofs}{{\indent \it \textbf{Proof}:\quad}}{\hfill $\blacksquare$\par}
\makeatother

\begin{document}
\title{On Convergence of Discrete Schemes for Computing the Rate-Distortion Function of Continuous Source
\thanks{$\dag$ Corresponding author.}
\thanks{This work was partially supported by National Key Research and Development Program of China (2018YFA0701603) and National Natural Science Foundation of China Grant (12271289 and 62231022).}
}

\author[1]{Lingyi Chen}
\author[1]{Shitong Wu}
\author[3$\dag$]{Wenyi Zhang}
\author[2]{Huihui Wu}
\author[1]{Hao Wu}
\affil[1]{Department of Mathematical Sciences, Tsinghua University, Beijing 100084, China}
\affil[2]{Yangtze Delta Region Institute (Huzhou), 
\authorcr University of Electronic Science and Technology of China, Huzhou, Zhejiang, 313000, P.R. China.}
\affil[3]{Department of Electronic Engineering and Information Science, 
\authorcr University of Science and Technology of China, Hefei, Anhui 230027, China 
\authorcr Email: wenyizha@ustc.edu.cn
}
\maketitle
\begin{abstract}
Computing the rate-distortion function for continuous sources is commonly regarded as a standard continuous optimization problem. When numerically addressing this problem, a typical approach involves discretizing the source space and subsequently solving the associated discrete problem. However, existing literature has predominantly concentrated on the convergence analysis of solving discrete problems, usually neglecting the convergence relationship between the original continuous optimization and its associated discrete counterpart. This neglect is not rigorous, since the solution of a discrete problem does not necessarily imply convergence to the solution of the original continuous problem, especially for non-linear problems. To address this gap, our study employs rigorous mathematical analysis, which constructs a series of finite-dimensional spaces approximating the infinite-dimensional space of the probability measure, establishing that solutions from discrete schemes converge to those from the continuous problems. 
\end{abstract}

\section{Introduction} \label{sec_introduction}
The lossy source coding theory \cite{book_element} and the rate-distortion function (RDF) for discrete memoryless sources were first established by Shannon \cite{shannon1948mathematical, shannon1959coding}. 
Until the end of 1960s, major developments of the classical setup had been summarized in \cite{berger1971}, including the lossy source coding theorem and associated RDF for sources and reproductions defined in abstract spaces.
The RDF that characterizes the fundamental tradeoff between compression rate and distortion, and plays a fundamental role in information theory. 
Several recent problems, including the rate-distortion-perception (RDP) function \cite{blau2019rethinking} and the information bottleneck (IB) \cite{tishby2000information_bottle}, have been formulated based on the concept of RDF. 
The RDF, along with its inspired extensions, has found extensive applications in fields such as image and video compression \cite{model_skodras2001jpeg,wiegand2003_H264} as well as machine learning \cite{balle2017end, alemi2016deep} problems.

The computation of RDF for discrete source and reproduction alphabets can be solved by the well-known Blahut-Arimoto (BA) algorithm \cite{blahut1972computation,arimoto1972algorithm}, which has been widely applied due to its conceptual simplicity. 
Recently, an intriguing connection has been identified between optimal transport (OT) and rate-distortion (RD) \cite{wu2022communication,yang2023estimating}. Leveraging insights from the OT perspective, more efficient algorithms have been developed for the computation of RDF; see, \textit{e.g.}, \cite{chen2023constrained, hayashi2023bregman}. 

For RD problems where the source alphabet is continuous, these algorithms are not directly applicable, and a general approach is to first discretize the continuous source space, and then solve the resulting discrete RD problem.
The solution of the discrete RD problem can be ensured by these algorithms.
However, one issue still remains and has usually been neglected: do the solutions of the discrete problems actually converge to those of the original continuous problems? 
To the best of our knowledge, we have not seen discussions of this issue in the relevant literature on the RD theory and related topics.
From a mathematical perspective, the continuous RD problem is a non-linear optimization problem. 
As is well-known, non-linear problems do not necessarily yield the solutions of the original continuous problems after discretization \cite{thomas2013numerical}.  
Moreover, similar discussion is also presented for OT problems, and the convergence of solutions for discrete problems to those for continuous problems \cite{Filippo2015optimal,graf2007foundations} is also specialized.
Hence, it is necessary to give a rigorous analysis for this issue.

Motivated by the discussion above, we propose a general framework for continuous RD problems and establish the convergence of their associated discrete problems.
The main idea in our proof is considering a proper form of the RDF and constructing a series of finite-dimensional spaces to approximate the infinite-dimensional space of the probability measure for the reproduction. 
Leveraging functional space analysis and estimation techniques \cite{rudin1976principles} and inspired by \cite{schochetman2001finite}, we rigorously prove the desired convergence property.
Notably, this convergence result is independent of specific numerical algorithms for discrete problems.

Furthermore, based on our framework, we provide a proof that $O(\frac{m|\log\varepsilon|}{\varepsilon^{d+1}})$ arithmetic operations are needed to achieve $\varepsilon$ accuracy using the BA algorithm \cite{blahut1972computation,arimoto1972algorithm} and that $O\big(\frac{m|\log\varepsilon|}{\varepsilon^{d+1}}(1+\log|\log\varepsilon|)\big)$ arithmetic operations while using the recently proposed Constrained BA (CBA) algorithm \cite{chen2023constrained}, which solves the RDF directly rather than the RD Lagrangian in the BA algorithm.
Here $m$ is the number of discretized nodes of the source $X$ when conducting numerical integration and $d$ is the dimension of the reproduction space $\mathcal{Y}$. 
It is worth mentioning that our framework may also be applied to other information theory problems including the IB problem and the RDP problem, due to their close connections in discrete formats.

For continuous RD problems, there is a considerable interest in determining the support set of reproduction, since in general, the optimal reproduction of a continuous source, under a mean squared error distortion, is discrete \cite{fix1977rate,fix1978rate,rose94}. 
To address this issue, a mapping approach has been proposed \cite{rose94} to optimize both the locations and the probability masses of the reproduction symbols. 
However, as remarked in \cite{mao2011rate}, there has not been theoretical guarantee for the mapping approach because taking into account the locations of reproduction symbols results in a non-convex optimization problem. 
In contrast, our method considers a series of discrete problems with their discretization step tending to zero, and the convergence is guaranteed since this does not involve determining the locations of reproduction symbols. 
While our convergence proof focuses on uniform grids of discretization, it is noteworthy that, with some technical modifications \cite{brenner2008mathematical}, our approach remains applicable to non-uniform grids.
This adaptability is crucial for designing more efficient discretization schemes, and is left for future research.

The remaining of this article is organised as follows. In Section II, the continuous and discrete RD problems are introduced and some mathematical notations are given. In Section III, the main theorem showing the convergence of discrete schemes is established. Then, in Section IV, the convergence rate and the complexity estimation are further provided. In Section V, numerical experiments are given to corroborate the analysis. The conclusion is drawn in Section VI.  

%%%%%%%%%%%%%%%%%%%%%%%%%%%%%%%%%%%%%%%%%%%%%%%%%%%%%%%%%%%%%%%%%%%%%%%%%%%%%%%%%%%%%%%%%%%%%%%%%%%%%%%%%%%%%%%%%%%%%%%%%%%%%%%%%%%%%%%%%%%%

\section{Preliminaries}

\subsection{The Continuous RD Problem}
Consider a continuous memoryless source $X\in\mathcal{X}$ with reproduction $Y\in\mathcal{Y}$. 
Here, $\mathcal{X},\mathcal{Y}$ are finite-dimensional Euclidean spaces. 
We denote the dimension of $\mathcal{Y}$ as $d$. 
In the sequel, denote the probability distributions of $X,Y$ as $p$, $\bdr$ respectively and the distortion measure as $\rho(x,y)$.
The original RD problem can be expressed as a max-min form (see \cite{lei2023neural}, equation (7)):
\begin{equation}\label{RD1}
        \max_{\beta\geq 0}\min_{\bm{r}\in W}-\int  \log[\int \exp(-\beta \rho(x,y))d\bdr(y)]dp(x)-\beta D,
    \end{equation}
    where $W=\{\bm{r}:\int d\bdr(y) =1\}$ and $D$ is the average distortion threshold. In the sequel, we denote the objective function in \eqref{RD1} as $ F(\bm{r},\beta)$. Here, $\bdr$ should be understood as a distribution and the topology on $W$ is the weak convergence topology \cite{munkres1999topology}.

By fixing the multiplier $\beta$, we have another form of the RD function (see \cite{rose94}, equation (3)), which has been used to develop the BA algorithm \cite{arimoto1972algorithm,blahut1972computation}
    \begin{equation}\label{RD}
        \min_{\bm{r}\in W} f(\bm{r})\triangleq-\int  \log[\int \exp(-\beta \rho(x,y))d\bdr(y)]dp(x).
    \end{equation}

\subsection{Discretization of the Continuous RD Problem}\label{sec_2_B}
Since $\bdr(y)$ is a continuous distribution, we discretize it at discrete points $\{y_j^n\}_{j=1}^n$, the equidistant nodes of $[-n^{\frac{1}{2d}},n^{\frac{1}{2d}}]^d$. We denote the discretization step size as $h$, \textit{i.e.}, $h=2 n^{-\frac{1}{2d}}$ and the interval $[y_j^n-h/2,y_j^n+h/2]$ containing $y_j$ as $I_j$ and note that these intervals are disjoint. For \eqref{RD}, the following discrete form can be given:
    \begin{equation}\label{RDd}
        \min_{\bm{r}:\sum_{j=1}^n r_j=1} -\int  \log[\sum_{j=1}^n \exp(-\beta \rho(x,y_j^n))r_j]dp(x).
    \end{equation}
%  %

 Correspondingly, \eqref{RD1} has the following discrete form:
 \begin{equation}\label{RD1d}
        \max_{\beta\geq 0}\min_{\bm{r}:\sum_{j=1}^n r_j=1}\!\! -\!\!\int\!  \log[\sum_{j=1}^n \exp(-\beta \rho(x,y_j^n))r_j]dp(x)-\beta D.
    \end{equation}
    We can express their solution as $\bm{r}^n=\sum_j r_j \delta_{y_j^n}$, where $\delta_{y_j^n}$ is the $\delta$ distribution at $y_j^n$.

\subsection{Notations from Mathematical Analysis}

In this paper, $\|\cdot\|$ represents the infinity norm and $\rightarrow$ means convergence in $\mathbb{R}$, while $\rightrightarrows$ means uniform convergence \cite{rudin1976principles}, \textit{i.e.}, $f_n\!\rightrightarrows\! f$ means that $\forall \varepsilon>0$, $|f_n(x)-f(x)|\!\leq\! \varepsilon,\forall x$, when $n$ is large enough.
We also denote the set that contains all the limit points of the sequence $\{\bm{r}^n\}_{n=1}^{\infty}$ as $L( \{\bm{r}^n\}_{n=1}^{\infty})$. The limit points \cite{munkres1999topology} of $\{\bm{r}^n\}_{n=1}^{\infty}$ are all the points satisfying that every punctured neighbourhood contains at least one point in $\{\bm{r}^n\}_{n=1}^{\infty}$.
Moreover, $[-M,M]^d$ means the Cartesian product $[-M,M]\times [-M,M]\times\cdots [-M,M]$, where $d$ intervals are involved. In the sequel, $\log\int \exp(-\beta \rho(x,y))d\bdr(y)$ appears frequently and for simplicity, we denote it as $\mathcal{H}_{\bdr(y)}(x)$.

%%%%%%%%%%%%%%%%%%%%%%%%%%%%%%%%%%%%%%%%%%%%%%%%%%%%%%%%%%%%%%
%%%%%%%%%%%%%%%%%%%%%%%%%%%%%%%%%%%%%%%%%%%%%%%%%%%%%%%%%%%%%%

\section{Convergence on Discrete Schemes} \label{sec_3}
To prove the convergence of discrete schemes, we need some mild assumptions:
 \begin{subequations}\label{assump}
           \begin{align}
               &\forall \beta>0,\forall \varepsilon>0,\exists \delta>0,\textit{ s.t. }\forall x,\forall\|y_1-y_2\|<\delta,
\quad\bigg|\exp(-\beta \rho(x,y_1))-\exp(-\beta \rho(x,y_2))\bigg|<\varepsilon,\label{assump_a}\\
     &\int \max_{\|y\|\leq A}\rho(x,y)dp(x)<\infty,\forall A>0,\label{assump_b}
           \end{align}
            \end{subequations}
    \begin{equation}\label{assump1}
            \begin{aligned}
            &\forall \beta>0,\forall \varepsilon>0,\exists \delta>0,\textit{ s.t. }\forall x,\forall\|y_1-y_2\|<\delta,\\
            &\bigg|\exp(-\beta \rho(x,y_1))\rho(x,y_1)-\exp(-\beta \rho(x,y_2))\rho(x,y_2)\bigg|<\varepsilon.
        \end{aligned}
        \end{equation}
            These are reasonable assumptions since for most cases including the mean absolute and mean squared error distortions, they hold.
\begin{theorems}\label{theorem1}
    Under Assumptions \eqref{assump}, 
    the solutions $\bm{r}^n$ to the discrete problem \eqref{RDd} satisfy both value convergence and sequence convergence, \textit{i.e.},
    \begin{equation*}
        f(\bm{r}^n)\rightarrow f^*, \text{ and } L(\{\bm{r}^n\}_{n=1}^{\infty}) \text{ is the solution set of \eqref{RD}},
    \end{equation*}
    where $f^*$ is the optimal value of the continuous problem \eqref{RD}.

\end{theorems}
\begin{proofs}
    We denote the actual optimal solution as $\bm{r}^*$ and $W_n=\{\sum_{j=1}^n r_j \delta_{y_j^n}| r_j\geq 0,\sum_{j=1}^n r_j= 1\}$. As shown in Section \ref{sec_2_B}, the continuous RD problem has the discrete form \eqref{RDd}, which is in fact minimizing $f(\bm{r})$ as defined in \eqref{RD}, over $W_n$.
    Denote the optimal solution of the discrete problem as $\bm{r}^n$.
    Then $f(\bm{r}^n)\leq f(\bm{r}), \forall \bm{r}\in W_n$.
    We observe that if we can find $\tilde{\bm{q}}^n\in W_n$ satisfying $f(\tilde{\bm{q}}^n)\rightarrow f(\bm{r}^*)$, then we have
    \begin{equation}\label{f}
        f(\bm{r}^*)\leq f(\bm{r}^n)\leq f(\tilde{\bm{q}}^n)\rightarrow f(\bm{r}^*).
    \end{equation}
    The first inequality is due to the fact $W_n\subseteq W$.
    The relationship \eqref{f} would then lead to the convergence $f(\bm{r}^n)\rightarrow f(\bm{r}^*)$.
    Now, we let $$\bm{q}^n=\sum_{j=1}^n A_j^n \delta_{y_j^n},$$ where $A_j^n=\int_{I_j}d\bm{r}^*$ and 
    $I_j$ is defined in Section \ref{sec_2_B}. Let $$\tilde{\bm{q}}^n=\bm{q}^n\bigg/\int d\bdq^n.$$ Since 
    $$\int d\bdq^n=\sum_j \int_{I_j}d\bm{r}^*\leq 1,$$ we have
\begin{equation*}
f(\tilde{\bm{q}}^n)=f(\bm{q}^n)+\log(\int d\bdq^n)\leq f(\bm{q}^n).
\end{equation*}
     Since $\tilde{\bm{q}}^n\in W_n$ and $f(\tilde{\bm{q}}^n)\leq f(\bm{q}^n)$, 
     we only need to prove $f({\bm{q}}^n)\rightarrow f(\bm{r}^*)$.
     Before the proof of $f({\bm{q}}^n)\rightarrow f(\bm{r}^*)$, we need to show the convergence after the truncation of $\mathcal{X}$ first.
        We let $A>1$ satisfy 
        $$\int_{\|y\|\leq A-1}d\bdr^*(y) \geq 1/2,$$
        and $M\geq A$ be given.
We first do the following estimation for $x\in [-M,M]^d$

    \begin{equation*}
        \begin{aligned}
            &\bigg|\int \exp(-\beta \rho(x,y))d\bdq^n(y)- \int \exp(-\beta \rho(x,y))d\bdr^*(y)\bigg|\\
            &\leq \sum_i\int_{I_i} \big|\exp(-\beta \rho(x,y))-\exp(-\beta \rho(x,y_i^n))\big|d\bdr^*(y)
            +\int_{|y|>n^{\frac{1}{2d}}}\exp(-\beta \rho(x,y))d\bdr^*(y)\\
            &\leq \sum_i\int_{I_i} \varepsilon d\bdr^*(y)+\int_{|y|>n^{\frac{1}{2d}}}d\bdr^*(y)\\
            &\leq 2\varepsilon,\text{ for sufficiently large $n$.}
        \end{aligned}
    \end{equation*}
    Here, we have used Assumptions \eqref{assump} in the second inequality and used the fact $$\bdq^n=\sum_j\int_{I_j}d\bdr^*\ \delta_{y_j^n}.$$
    Next we establish the uniform convergence.
    Note that for $x\in[-M,M]^d$,
    \begin{equation}\label{lower_bound}
        \begin{aligned}
            &\int \exp(-\beta \rho(x,y))d\bdq^n(y)\geq \int_{\|y\|\leq M} \exp(-\beta \rho(x,y))d\bdq^n(y)\\
            &\geq \int_{\|y\|\leq M} \exp(-\beta \rho^*)d\bdq^n(y)\geq \frac{1}{2} \exp(-\beta \rho^*)\triangleq \delta.
        \end{aligned}
    \end{equation}
    Here $\rho^*=\max_{x,y\in[-M,M]^d}\rho(x,y)$ and $\int_{\|y\|\leq M} d\bdq^n\geq \frac{1}{2}$ is due to 
    \begin{equation}\label{1/2}
    \begin{aligned}
        \int_{\|y\|\leq A} d\bdq^n(y)&=\sum_{i:y_i^n\in [-A,A]^d} \int_{I_i} d\bdr^*(y)
        \geq \int_{\|y\|\leq A-1}d\bdr^*(y)\geq \frac{1}{2} .
    \end{aligned}
    \end{equation}
    Since $\log x$ is uniformly continuous in $[\delta,+\infty)$ and $\int \exp(-\beta \rho(x,y))d\bdq^n(y)$ uniformly converges, we have 
        \begin{multline*}
            \log\bigg(\int \exp(-\beta \rho(x,y))d\bdq^n(y)\bigg)\rightrightarrows
            \log\bigg(\int \exp(-\beta \rho(x,y))d\bdr^*(y)\bigg),\quad x\in [-M,M]^d.
      \end{multline*}
    Then 
    \begin{equation*}
        \begin{aligned}
            &\bigg|\int_{\|x\|\leq M}  \log\bigg[\int \exp(-\beta \rho(x,y))d\bdq^n(y)\bigg]dp(x)
            - \int_{\|x\|\leq M}  \log\bigg[\int \exp(-\beta \rho(x,y))d\bdr^*(y)\bigg]dp(x)\bigg|\\
            &\leq \int_{\|x\|\leq M} \varepsilon dp(x) \leq \varepsilon,\text{ when $n$ is sufficiently large}.
        \end{aligned}
    \end{equation*}        
        
        Then as $n\rightarrow+\infty$, we have 
        \begin{multline*}
            \int_{\|x\|\leq M}  \log\bigg[\int \exp(-\beta \rho(x,y))d\bdq^n(y)\bigg]dp(x)\rightarrow
            \int_{\|x\|\leq M}  \log\bigg[\int \exp(-\beta \rho(x,y))d\bdr^*(y)\bigg]dp(x).
        \end{multline*}
    Thus, we obtain the convergence after truncation. In the following, based on this result, we prove the convergence $f({\bm{q}}^n)\rightarrow f(\bm{r}^*)$. For $M\geq A$,
     \begin{equation}\label{err}
        \begin{aligned}
            &|f({\bm{q}}^n)-f(\bm{r}^*)|=\bigg|\int  \mathcal{H}_{{\bm{q}}^n(y)}(x)dp(x)-\int  \mathcal{H}_{\bm{r}^*(y)}(x)dp(x)\bigg|\\
            &\leq \bigg|\int_{\|x\|\leq M}  [\mathcal{H}_{{\bm{q}}^n(y)}(x)-\mathcal{H}_{\bm{r}^*(y)}(x)]dp(x)\bigg|
            +\int_{\|x\|>M}  \bigg(-\mathcal{H}_{{\bm{q}}^n(y)}(x)-\mathcal{H}_{\bm{r}^*(y)}(x)\bigg)dp(x),\\
        \end{aligned}
     \end{equation} 
     where $$\mathcal{H}_{\bdr(y)}(x)=\log\int \exp(-\beta \rho(x,y))d\bdr(y).$$
     We denote $g(x)=\max_{\|y\|\leq A}\rho(x,y)$, and note that
     \begin{equation*}
        \begin{aligned}
            \int \exp(-\beta \rho(x,y))d\bdq^n(y)&\geq \exp(-\beta g(x))\int_{\|y\|\leq A} d\bdq^n(y)
            \geq \frac{1}{2} \exp(-\beta g(x)).
        \end{aligned}
     \end{equation*}
     Here the second inequality is similar to \eqref{1/2}.
    Similarly, 
    $$\int \exp(-\beta \rho(x,y))d\bdr^*(y)\geq \frac{1}{2} \exp(-\beta g(x)).$$
    We thus obtain the following evaluation
    \begin{equation*}
        \begin{aligned}
            &\int_{\|x\|>M}  \bigg(-\mathcal{H}_{{\bm{q}}^n}(x)-\mathcal{H}_{\bm{r}^*}(x)\bigg)dp(x)
            \leq \int_{\|x\|>M}(2\log2+2\beta g(x))dp(x)\rightarrow 0,\text{ as }M\rightarrow +\infty.
        \end{aligned}
    \end{equation*}
    Here Assumption \eqref{assump_b}, \textit{i.e.}, $\int g(x)dp(x)<\infty$ has been used.
    Thus, by \eqref{err}, we have 
   \begin{equation*}
        \begin{aligned}
            |f({\bm{q}}^n)-f(\bm{r}^*)|
            &\leq \bigg|\int_{\|x\|\leq M}  [\mathcal{H}_{{\bm{q}}^n(y)}(x)-\mathcal{H}_{\bm{r}^*(y)}(x)]dp(x)\bigg|
            +\int_{\|x\|>M}(2\log2+2\beta g(x))dp(x).
        \end{aligned}
     \end{equation*} 
    By taking the upper limit of $n$ for this inequality, we obtain
    \begin{equation*}
        \begin{aligned}
            &\limsup_n \bigg|f({\bm{q}}^n)-f(\bm{r}^*)\bigg|
            \leq \int_{\|x\|>M}(2\log2+2\beta g(x))dp(x).\\
        \end{aligned}
    \end{equation*}
    By letting $M\rightarrow+\infty$, we obtain $f(\bm{q}^n)\rightarrow f(\bm{r}^*)$.\\
    To prove the convergence of the solutions, we let $\tilde{\bm{r}}$ be a limit point of the solution sequence 
    $\{\bm{r}^n\}_{n=1}^{\infty}$. 
    Then there exists a subsequence
    $\{\bm{r}^{n_k}\}_{k=1}^{\infty}$ satisfying 
    $\bm{r}^{n_k}\rightarrow \tilde{\bm{r}}$. Next, we have 
    $$f(\tilde{\bm{r}})=\lim_k f(\bm{r}^{n_k})=f(\bm{r}^*),$$
    since we have proven $\lim_n f(\bm{r}^{n})=f(\bm{r}^*)$.
    Therefore, we have shown that $f(\tilde{\bm{r}})$ is equal to the optimal value, and consequently $\tilde{\bm{r}}$ is an optimal solution.
    
\end{proofs}

Next, we provide the convergence proof of the RD problem \eqref{RD1}. Due to its max-min form and additional variable $\beta$, the proof is more complicated than that of Theorem \ref{theorem1}.

\begin{theorems}\label{theorem2}
    Under Assumptions \eqref{assump} and \eqref{assump1}, the solutions $(\bm{r}^n,\beta^n)$ to the discrete problem \eqref{RD1d} satisfy both value convergence and sequence convergence, \textit{i.e.},
    \begin{equation*}
    F(\bm{r}^n,\beta^n)\rightarrow F^*, \text{ and } L  (\{(\bm{r}^{n},\beta^{n})\}_{n=1}^{\infty}) \text{ are solutions of \eqref{RD1}},
    \end{equation*}
    where $F^*$ is the optimal value of the continuous problem \eqref{RD1}.
    \end{theorems}
\begin{proofs}
    We denote the optimal solution of the continuous problem \eqref{RD1} and the discrete problem \eqref{RD1d} as $(\bm{r}^*,\beta^*),(\bm{r}^n,\beta^n)$ respectively. Denote $W_n=\{\sum_{j=1}^n c_j \delta_{y_j^n}| c_j\geq 0,\sum_{j=1}^n c_j= 1\}$, where $\delta_{y_j^n}$ is the $\delta$ distribution at $y_j^n$. Similar to Theorem \ref{theorem1}, the discrete problem \eqref{RD1d} is equivalent to optimizing $F(\bm{r},\beta)$ over $W_n$.
     By the property of max-min problems \eqref{RD1d} and \eqref{RD1}, we have
    $\forall \bdr\in W_n,\forall \beta\geq 0$,
    $$F(\bdr^n,\beta^n)\leq F(\bdr,\beta^n),\ F(\bdr^n,\beta^n)\geq F(\bdr^n,\beta).$$
    Furthermore, $\forall \bdr\in W,\forall \beta\geq 0$,
    $$F(\bdr^*,\beta^*)\leq F(\bdr,\beta^*),\ F(\bdr^*,\beta^*)\geq F(\bdr^*,\beta).$$
    Next, we construct $\bm{q}^n=\sum_{j=1}^n A_j^n \delta_{y_j^n}$, here $$A_j^n=\int_{I_j}d\bm{r}^*,$$
    $I_j$ is the interval $[y_j^n-h/2,y_j^n+h/2]$ containing $y_j$. Let $$\tilde{\bm{q}}^n=\bm{q}^n\bigg/\int d\bdq^n \in W_n.$$ 
    Since $\int d\bdq^n\leq 1$, we have
$$F(\tilde{\bm{q}}^n,\beta)=F(\bm{q}^n,\beta)+\log(\int d\bdq^n)\leq F(\bm{q}^n,\beta),\quad\forall \beta\geq 0.$$
    Then we have the following chain of inequalities:
    \begin{equation*}
        F(\bm{r}^*,\beta^*)\leq F(\bdr^n,\beta^*)\leq F(\bdr^n,\beta^n)\leq F(\tilde{q}^n,\beta^n)\leq F(\bdq^n,\beta^n)\leq F(\bdq^n,\tilde{\beta}^n),
    \end{equation*}
    where $\tilde{\beta}^n=\text{argmax}_{\beta} F(\bdq^n,\beta)$.
    We observe that if $F(\bm{q}^n,\tilde{\beta}^n)\rightarrow F(\bm{r}^*,\beta^*)$, then we have
    \begin{equation}\label{f1}
        F(\bm{r}^*,\beta^*)\leq F(\bm{r}^n,\beta^n)\leq F(\bm{q}^n,\tilde{\beta}^n)\rightarrow F(\bm{r}^*,\beta^*).
    \end{equation}
    Therefore, we obtain the convergence $F(\bm{r}^n,\beta^n)\rightarrow F(\bm{r}^*,\beta^*)$.
    So in summary, we only need to prove $F(\bm{q}^n,\tilde{\beta}^n)\rightarrow F(\bm{r}^*,\beta^*)$.
    
    From the optimality of $\tilde{\beta}^n$, we know $\tilde{\beta}^n$ should satisfy the first order condition and is the root of a monotone function 
    \begin{equation*}
        G_{\bdq^n}(\beta)=\int \bigg(\int e^{-\beta \rho(x,y)}\rho(x,y)d\bdq^n(y)\bigg)\bigg/\bigg(\int e^{-\beta \rho(x,y)}d\bdq^n(y)\bigg) dp(x)-D.
    \end{equation*}
    The monotone property is due to Cauchy inequality
    \begin{equation*}
         G_{\bdq^n}^{\prime}(\beta)=-\int\frac{\bigg[\bigg(\int e^{-\beta \rho(x,y)}d\bdq^n(y)\bigg)\bigg(\int e^{-\beta \rho(x,y)}\rho(x,y)^2d\bdq^n(y)\bigg) -\bigg(\int e^{-\beta \rho(x,y)}\rho(x,y)d\bdq^n(y)\bigg)^2\bigg]}{\bigg(\int e^{-\beta \rho(x,y)}d\bdq^n(y)\bigg)^2} dp(x)< 0.
    \end{equation*}
    It does not equal to 0, otherwise, by the condition of equality for Cauchy inequality, $\rho(x,y)$ is a function only with respect to $x$, and this degenerate case can be disregarded.
    
    \textbf{We first show $\tilde{\beta}^n$ is lower bounded for $n$, \textit{i.e.}, $\tilde{\beta}^n\geq B_1>0,\forall n$}. Otherwise, there exists a subsequence $\tilde{\beta}^{n_k}\rightarrow 0$. We will show it is a contradiction. 
    Since $\tilde{\beta}^{n_k}\rightarrow 0$, we have 
    \begin{equation*}
        \exists k_0,\forall k>k_0,\ \tilde{\beta}^{n_k}<\beta^*/2.
    \end{equation*}
    Here, we assume $\beta^*>0$ and the degenerate case $\beta^*=0$ can be disregarded, in which the rate equals to 0.
    Then by the monotone property of $G_{\bdq^{n_k}}$, we have 
    \begin{equation*}
        0=G_{\bdq^{n_k}}(\tilde{\beta}^{n_k})>G_{\bdq^{n_k}}(\beta^*/2).
    \end{equation*}
    Thus,
    \begin{equation*}
        \begin{aligned}
            D&> \int \bigg(\int e^{-\beta^* \rho(x,y)/2}\rho(x,y)d\bdq^{n_k}(y)\bigg)\bigg/\bigg(\int e^{-\beta^* \rho(x,y)/2}d\bdq^{n_k}(y)\bigg) dp(x)\\
            &\geq \int_{\|x\|\leq M} \bigg(\int e^{-\beta^* \rho(x,y)/2}\rho(x,y)d\bdq^{n_k}(y)\bigg)\bigg/\bigg(\int e^{-\beta^* \rho(x,y)/2}d\bdq^{n_k}(y)\bigg) dp(x).\\
        \end{aligned}
    \end{equation*}
    Here, $M$ is a sufficiently large parameter.
    By the proof of Theorem \ref{theorem1}, we have 
    \begin{equation*}
        \int e^{-\beta^* \rho(x,y)/2}d\bdq^{n_k}(y)\rightrightarrows \int e^{-\beta^* \rho(x,y)/2}d\bdr^*(y).
    \end{equation*}
    Similarly, 
    \begin{equation*}
        \int e^{-\beta^* \rho(x,y)/2}\rho(x,y)d\bdq^{n_k}(y)\rightrightarrows \int e^{-\beta^* \rho(x,y)/2}\rho(x,y)d\bdr^*(y).
    \end{equation*}
    And similar to \eqref{lower_bound} in the proof of Theorem \ref{theorem1}, we have
    \begin{equation*}
        \int e^{-\beta^* \rho(x,y)/2}d\bdq^{n_k}(y)\geq \delta_0.
    \end{equation*}
    Meanwhile, we have an upper bound estimation
    \begin{equation*}
        \int e^{-\beta^* \rho(x,y)/2}\rho(x,y)d\bdq^{n_k}(y)\leq \int B\ d\bdq^{n_k}(y)\leq B.
    \end{equation*}
    Here, $B$ is the bound of the function $e^{-\beta^* x/2}x,x\geq0$.
    Next, since $g(x,y)=x/y$ is uniformly continuous in $[0,B]\times [\delta_0,\infty)$, we obtain
    \begin{equation*}
        \frac{\bigg(\int \exp({-\beta^* \rho(x,y)/2})\rho(x,y)d\bdq^{n_k}(y)\bigg)}{\bigg(\int \exp({-\beta^* \rho(x,y)/2})d\bdq^{n_k}(y)\bigg)}\rightrightarrows \frac{\bigg(\int \exp({-\beta^* \rho(x,y)/2})\rho(x,y)d\bdr^*(y)\bigg)}{\bigg(\int \exp({-\beta^* \rho(x,y)/2})d\bdr^*(y)\bigg)}
    \end{equation*}
    Thus,
    \begin{equation}\label{conv1}
        \begin{aligned}
            D&\geq \int_{\|x\|\leq M} \bigg(\int e^{-\beta^* \rho(x,y)/2}\rho(x,y)d\bdq^{n_k}(y)\bigg)\bigg/\bigg(\int e^{-\beta^* \rho(x,y)/2}d\bdq^{n_k}(y)\bigg) dp(x)\\
            &\quad \rightarrow \int_{\|x\|\leq M} \bigg(\int e^{-\beta^* \rho(x,y)/2}\rho(x,y)d\bdr^*(y)\bigg)\bigg/\bigg(\int e^{-\beta^* \rho(x,y)/2}d\bdr^*(y)\bigg) dp(x)
        \end{aligned}
    \end{equation}
We let $M\rightarrow \infty$, and get
\begin{equation*}
    G_{\bdr^*}(\beta^*/2)\leq 0=G_{\bdr^*}(\beta^*).
\end{equation*}
Using the monotonicity of $G_{\bdr^*}$, we have $\beta^*/2\geq \beta^*$, which is a contradiction.

\textbf{Next, we will prove the convergence of $\tilde{\beta}^n$.}
To prove $\tilde{\beta}^n\rightarrow \beta^*$, we only need to show every convergent subsequence converges to $\beta^*$, \textit{i.e.}, if a convergent subsequence $\tilde{\beta}^{n_k}\rightarrow \Bar{\beta}$, then $\Bar{\beta}=\beta^*$. Now, let $\tilde{\beta}^{n_k}$ be a convergent subsequence and $\tilde{\beta}^{n_k}\rightarrow\Bar{\beta}$. Since $\tilde{\beta}^{n_k}$ is lower bounded, we have $\Bar{\beta}>0$.
Let $A_0$ satisfies $e^{-\Bar{\beta}t/2}t\leq \varepsilon$ and $e^{-\Bar{\beta}t/2}\leq \varepsilon,\ \forall t\geq A_0$.
\begin{equation*}
    \begin{aligned}
        &\bigg|\int e^{-\tilde{\beta}^{n_k}\rho(x,y)}\rho(x,y)d\bdq^{n_k}(y)-\int e^{-\Bar{\beta}\rho(x,y)}\rho(x,y)d\bdq^{n_k}(y)\bigg|\leq \int |e^{-\tilde{\beta}^{n_k}\rho(x,y)}-e^{-\Bar{\beta}\rho(x,y)}|\rho(x,y)d\bdq^{n_k}(y)\\
        &\leq \int e^{-\Bar{\beta}\rho(x,y)}|e^{(\Bar{\beta}-\tilde{\beta}^{n_k})\rho(x,y)}-1|\rho(x,y)d\bdq^{n_k}(y)\\
    \end{aligned}
\end{equation*}
Next, we divide it into two parts and estimate each part.
\begin{equation*}
    \begin{aligned}
        &\int_{y:\rho(x,y)>A_0} e^{-\Bar{\beta}\rho(x,y)}|e^{(\Bar{\beta}-\tilde{\beta}^{n_k})\rho(x,y)}-1|\rho(x,y)d\bdq^{n_k}(y)\\
        &\leq \int_{y:\rho(x,y)>A_0} e^{-\Bar{\beta}\rho(x,y)}e^{|\Bar{\beta}-\tilde{\beta}^{n_k}|\rho(x,y)}\rho(x,y)d\bdq^{n_k}(y)\\
        &\leq \int_{y:\rho(x,y)>A_0} e^{-\Bar{\beta}\rho(x,y)/2}\rho(x,y)d\bdq^{n_k}(y)\\
        &\leq \int_{y:\rho(x,y)>A_0} \varepsilon d\bdq^{n_k}(y)\leq \varepsilon, \text{ when k is sufficiently large}
    \end{aligned}
\end{equation*}
here, the second inequality is due to $|\Bar{\beta}-\tilde{\beta}^{n_k}|\leq \Bar{\beta}/2$, when k is sufficiently large.\\
\begin{equation*}
    \begin{aligned}
        &\int_{y:\rho(x,y)\leq A_0} e^{-\Bar{\beta}\rho(x,y)}|e^{(\Bar{\beta}-\tilde{\beta}^{n_k})\rho(x,y)}-1|\rho(x,y)d\bdq^{n_k}(y)\\
        &\leq\int_{y:\rho(x,y)\leq A_0} |e^{(\Bar{\beta}-\tilde{\beta}^{n_k})\rho(x,y)}-1|A_0\ d\bdq^{n_k}(y)\\
        &\leq\int_{y:\rho(x,y)\leq A_0} (e^{|\Bar{\beta}-\tilde{\beta}^{n_k}|\rho(x,y)}-1)A_0\ d\bdq^{n_k}(y)\\
        &\leq\int_{y:\rho(x,y)\leq A_0} (e^{|\Bar{\beta}-\tilde{\beta}^{n_k}|A_0}-1)A_0\ d\bdq^{n_k}(y)\\
        &\leq (e^{|\Bar{\beta}-\tilde{\beta}^{n_k}|A_0}-1)A_0\leq \varepsilon, \text{ when k is sufficiently large}
    \end{aligned}
\end{equation*}
Combining the two parts, we obtain
\begin{equation*}
    \bigg|\int e^{-\tilde{\beta}^{n_k}\rho(x,y)}\rho(x,y)d\bdq^{n_k}(y)-\int e^{-\Bar{\beta}\rho(x,y)}\rho(x,y)d\bdq^{n_k}(y)\bigg|\leq 2\varepsilon, \text{ when k is sufficiently large}.
\end{equation*}
Thus,
\begin{equation}\label{conv}
    \begin{aligned}
        \int e^{-\tilde{\beta}^{n_k}\rho(x,y)}\rho(x,y)d\bdq^{n_k}(y)-\int e^{-\Bar{\beta}\rho(x,y)}\rho(x,y)d\bdq^{n_k}(y)\rightrightarrows 0,\  \forall x.
    \end{aligned}
\end{equation}
And by the proof of Theorem \ref{theorem1}, we have
\begin{equation*}
    \begin{aligned}
        \int e^{-\Bar{\beta}\rho(x,y)}\rho(x,y)d\bdq^{n_k}(y)\rightrightarrows\int e^{-\Bar{\beta}\rho(x,y)}\rho(x,y)d\bdr^*(y),\  \forall x.
    \end{aligned}
\end{equation*}
Thus,
\begin{equation*}
    \begin{aligned}
        \int e^{-\tilde{\beta}^{n_k}\rho(x,y)}\rho(x,y)d\bdq^{n_k}(y)\rightrightarrows\int e^{-\Bar{\beta}\rho(x,y)}\rho(x,y)d\bdr^*(y),\  \forall x.
    \end{aligned}
\end{equation*}
Similarly, we have
\begin{equation*}
    \begin{aligned}
        \int e^{-\tilde{\beta}^{n_k}\rho(x,y)}d\bdq^{n_k}(y)\rightrightarrows\int e^{-\Bar{\beta}\rho(x,y)}d\bdr^*(y),\  \forall x.
    \end{aligned}
\end{equation*}
%%%%%%%%%%%
%
%
%
Next, we will prove $G_{\bdq^{n_k}}(\tilde{\beta}^{n_k})\rightarrow G_{\bdr^*}(\Bar{\beta})$.
Since $\tilde{\beta}^{n_k}\rightarrow \Bar{\beta}$, we have $|\tilde{\beta}^{n_k}- \Bar{\beta}|\leq \varepsilon$, when $k$ is sufficiently large. Using the monotonicity of $G_{\bdq^{n_k}}$, we obtain 
\begin{equation}\label{ineq}
   G_{\bdq^{n_k}}(\Bar{\beta}+\varepsilon) \leq G_{\bdq^{n_k}}(\tilde{\beta}^{n_k})\leq G_{\bdq^{n_k}}(\Bar{\beta}-\varepsilon).
\end{equation}
Next, We divide $G_{\bdq^{n_k}}(\Bar{\beta}+\varepsilon)$ into two parts, and estimate each part. For simplicity, we denote $\Bar{\beta}+\varepsilon$ as $\beta$. We first estimate the part
\begin{equation*}
    \begin{aligned}
        & \int_{\|x\|>M} \bigg(\int e^{-\beta \rho(x,y)}\rho(x,y)d\bdq^{n_k}(y)\bigg)\bigg/\bigg(\int e^{-\beta \rho(x,y)}d\bdq^{n_k}(y)\bigg) dp(x).\\
    \end{aligned}
\end{equation*}
We denote $A_x=-\frac{1}{\beta}\ \log \int e^{-\beta \rho(x,y)}d\bdq^{n_k}(y)$, then
\begin{equation*}
    \begin{aligned}
        &\int e^{-\beta \rho(x,y)}\rho(x,y)d\bdq^{n_k}(y)\\
        &\leq \int_{y: \rho(x,y)\leq A_x} e^{-\beta \rho(x,y)}\rho(x,y)d\bdq^{n_k}(y)+ \int_{y: \rho(x,y)> A_x} e^{-\beta \rho(x,y)}\rho(x,y)d\bdq^{n_k}(y)\\
        &\leq \int_{y: \rho(x,y)\leq A_x} e^{-\beta \rho(x,y)}A_x\ d\bdq^{n_k}(y)+ \int_{y: \rho(x,y)> A_x} e^{-\beta \max(A_x,1/\beta)}\max(A_x,1/\beta)d\bdq^{n_k}(y)\\
        &\leq A_x \int e^{-\beta \rho(x,y)}\ d\bdq^{n_k}(y)+ e^{-\beta \max(A_x,1/\beta)}\max(A_x,1/\beta)\\
        &=A_x \exp(-\beta A_x)+ e^{-\beta \max(A_x,1/\beta)}\max(A_x,1/\beta)
    \end{aligned}
\end{equation*}
Thus,
\begin{equation*}
    \begin{aligned}
        & \int_{\|x\|>M} \bigg(\int e^{-\beta \rho(x,y)}\rho(x,y)d\bdq^{n_k}(y)\bigg)\bigg/\bigg(\int e^{-\beta \rho(x,y)}d\bdq^{n_k}(y)\bigg) dp(x)\\
        &\leq \int_{\|x\|>M}\bigg(A_x \exp(-\beta A_x)+ e^{-\beta \max(A_x,1/\beta)}\max(A_x,1/\beta)\bigg)\bigg/ \exp(-\beta A_x) dp(x)\\
        &\leq \int_{\|x\|>M}\bigg(A_x+ \max(A_x,1/\beta)\bigg) dp(x)\\
        &\leq \int_{\|x\|>M}\bigg(2 A_x+ 1/\beta\bigg) dp(x)\\
    \end{aligned}
\end{equation*}
Note that
\begin{equation*}
    \begin{aligned}
        A_x&=-\frac{1}{\beta}\ \log \int e^{-\beta \rho(x,y)}d\bdq^{n_k}(y)\\
        &\leq -\frac{1}{\beta}\ \log \int_{\|y\|\leq A} e^{-\beta \rho(x,y)}d\bdq^{n_k}(y)\\
        &\leq -\frac{1}{\beta}\ \log \int_{\|y\|\leq A} e^{-\beta g(x)}d\bdq^{n_k}(y)\\
        &\leq -\frac{1}{\beta}\ \log(e^{-\beta g(x)}/2)\\
        &=g(x)+(\log 2)/\beta, \text{\ when k is sufficiently large.}
    \end{aligned}
\end{equation*}
Here, $A$ is the constant in Theorem \ref{theorem1} and $g(x)=\max_{\|y\|\leq A}\rho(x,y)$. And we get $$\int \bigg(2 A_x+ 1/\beta\bigg) dp(x)<\infty.$$
Similar to \eqref{conv1}, the other part
\begin{equation*}
    \begin{aligned}
        & \int_{\|x\|\leq M} \bigg(\int e^{-\beta \rho(x,y)}\rho(x,y)d\bdq^{n_k}(y)\bigg)\bigg/\bigg(\int e^{-\beta \rho(x,y)}d\bdq^{n_k}(y)\bigg) dp(x)\\
        &\rightarrow \int_{\|x\|\leq M} \bigg(\int e^{-\beta \rho(x,y)}\rho(x,y)d\bdr^*(y)\bigg)\bigg/\bigg(\int e^{-\beta \rho(x,y)}d\bdr^*(y)\bigg) dp(x), \text{as }k\rightarrow \infty.
    \end{aligned}
\end{equation*}
Finally, combining the estimate of the two parts, we have
\begin{equation*}
    \begin{aligned}
        &G_{\bdq^{n_k}}(\beta)=\int \bigg(\int e^{-\beta \rho(x,y)}\rho(x,y)d\bdq^{n_k}(y)\bigg)\bigg/\bigg(\int e^{-\beta \rho(x,y)}d\bdq^{n_k}(y)\bigg) dp(x)-D\\
        &\leq \int_{\|x\|\leq M} \bigg(\int e^{-\beta \rho(x,y)}\rho(x,y)d\bdq^{n_k}(y)\bigg)\bigg/\bigg(\int e^{-\beta \rho(x,y)}d\bdq^{n_k}(y)\bigg) dp(x)+\int_{\|x\|>M}\bigg(2 A_x+ 1/\beta\bigg) dp(x)-D\\
    \end{aligned}
\end{equation*}
We first let $k$ goes to $\infty$, 
\begin{equation*}
    \begin{aligned}
        \limsup_k G_{\bdq^{n_k}}(\beta)&\leq \int_{\|x\|\leq M} \bigg(\int e^{-\beta \rho(x,y)}\rho(x,y)d\bdr^*(y)\bigg)\bigg/\bigg(\int e^{-\beta \rho(x,y)}d\bdr^*(y)\bigg) dp(x)\\
        &\quad+\int_{\|x\|>M}\bigg(2 A_x+ 1/\beta\bigg) dp(x)-D\\
    \end{aligned}
\end{equation*}
Then let $M\rightarrow\infty$, we obtain
\begin{equation*}
    \begin{aligned}
        &\limsup_k G_{\bdq^{n_k}}(\beta)\leq \int \bigg(\int e^{-\beta \rho(x,y)}\rho(x,y)d\bdr^*(y)\bigg)\bigg/\bigg(\int e^{-\beta \rho(x,y)}d\bdr^*(y)\bigg) dp(x)-D=G_{\bdr^*}(\beta),\\
    \end{aligned}
\end{equation*}
\textit{i.e.}, $\limsup_k G_{\bdq^{n_k}}(\Bar{\beta}+\varepsilon)\leq G_{\bdr^*}(\Bar{\beta}+\varepsilon)$.\\
Meanwhile,
\begin{equation*}
    \begin{aligned}
        &G_{\bdq^{n_k}}(\beta)=\int \bigg(\int e^{-\beta \rho(x,y)}\rho(x,y)d\bdq^{n_k}(y)\bigg)\bigg/\bigg(\int e^{-\beta \rho(x,y)}d\bdq^{n_k}(y)\bigg) dp(x)-D\\
        &\geq \int_{\|x\|\leq M} \bigg(\int e^{-\beta \rho(x,y)}\rho(x,y)d\bdq^{n_k}(y)\bigg)\bigg/\bigg(\int e^{-\beta \rho(x,y)}d\bdq^{n_k}(y)\bigg) dp(x)-D\\
    \end{aligned}
\end{equation*}
Let $k\rightarrow\infty$ and then $M\rightarrow\infty$, we have
\begin{equation*}
    \begin{aligned}
        &\liminf_k G_{\bdq^{n_k}}(\beta)\geq \lim_{M\rightarrow\infty}\int_{\|x\|\leq M} \bigg(\int e^{-\beta \rho(x,y)}\rho(x,y)d\bdr^*(y)\bigg)\bigg/\bigg(\int e^{-\beta \rho(x,y)}d\bdr^*(y)\bigg) dp(x)-D=G_{\bdr^*}(\beta),\\
    \end{aligned}
\end{equation*}
\textit{i.e.}, $\liminf_k G_{\bdq^{n_k}}(\Bar{\beta}+\varepsilon)\geq G_{\bdr^*}(\Bar{\beta}+\varepsilon)$.
Thus, we obtain $$\lim_k G_{\bdq^{n_k}}(\Bar{\beta}+\varepsilon)= G_{\bdr^*}(\Bar{\beta}+\varepsilon).$$
Similarly, $$\lim_k G_{\bdq^{n_k}}(\Bar{\beta}-\varepsilon)= G_{\bdr^*}(\Bar{\beta}-\varepsilon).$$
Using \eqref{ineq}, we have
\begin{equation*}
  G_{\bdr^*}(\Bar{\beta}+\varepsilon)=  \lim_k G_{\bdq^{n_k}}(\Bar{\beta}+\varepsilon)\leq \liminf_k G_{\bdq^{n_k}}(\tilde{\beta}^{n_k})\leq \limsup_k G_{\bdq^{n_k}}(\tilde{\beta}^{n_k})\leq \lim_k G_{\bdq^{n_k}}(\Bar{\beta}-\varepsilon)=G_{\bdr^*}(\Bar{\beta}-\varepsilon).
\end{equation*}
Then, let $\varepsilon\rightarrow 0$, we obtain $0=\lim_k G_{\bdq^{n_k}}(\tilde{\beta}^{n_k})=G_{\bdr^*}(\Bar{\beta})$.
Therefore, $G_{\bdr^*}(\Bar{\beta})=0=G_{\bdr^*}(\beta^*)$. Then by the strict monotonicity, we have $\Bar{\beta}=\beta^*$. Thus $\tilde{\beta}^{n}\rightarrow \beta^*$ holds.

\textbf{Now, we will show the convergence $F(\bdq^n,\tilde{\beta}^{n})\rightarrow F(\bdr^*,\beta^*)$.} By Theorem 1, we have $F(\bdq^n,\beta^*)\rightarrow F(\bdr^*,\beta^*)$. Thus we only need to show $F(\bdq^n,\tilde{\beta}^{n})-F(\bdq^n,\beta^*)\rightarrow 0$.\\
\begin{equation*}
    \begin{aligned}
        0\leq F(\bdq^n,\tilde{\beta}^{n})-F(\bdq^n,\beta^*)&=-\int \log \frac{\int \exp(-\tilde{\beta}^{n}\rho(x,y))d\bdq^n(y)}{\int \exp(-\beta^*\rho(x,y))d\bdq^n(y)}dp(x)-(\tilde{\beta}^{n}-\beta^*)D\\
    \end{aligned}
\end{equation*}
We divide the integration into two parts and estimate each part.
\begin{equation*}
    \begin{aligned}
        &-\int_{\|x\|\leq M} \log \frac{\int \exp(-\tilde{\beta}^{n}\rho(x,y))d\bdq^n(y)}{\int \exp(-\beta^*\rho(x,y))d\bdq^n(y)}dp(x)\\
        &=\int_{\|x\|\leq M} \log \bigg(\frac{\int \exp(-\beta^*\rho(x,y))d\bdq^n(y)-\int \exp(-\tilde{\beta}^{n}\rho(x,y))d\bdq^n(y)}{\int \exp(-\tilde{\beta}^{n}\rho(x,y))d\bdq^n(y)}+1\bigg)dp(x)\\
        &\leq \int_{\|x\|\leq M}\frac{|\int \exp(-\beta^*\rho(x,y))d\bdq^n(y)-\int \exp(-\tilde{\beta}^{n}\rho(x,y))d\bdq^n(y)|}{\int \exp(-\tilde{\beta}^{n}\rho(x,y))d\bdq^n(y)}dp(x)\\
        &\leq \int_{\|x\|\leq M}\frac{|\int \exp(-\beta^*\rho(x,y))d\bdq^n(y)-\int \exp(-\tilde{\beta}^{n}\rho(x,y))d\bdq^n(y)|}{\delta_0}dp(x)
    \end{aligned}
\end{equation*}
Here, $\delta_0$ is a lower bound
\begin{equation*}
    \begin{aligned}
        \int \exp(-\tilde{\beta}^{n}\rho(x,y))d\bdq^n(y)&\geq \int_{\|y\|\leq M} \exp(-B_0 \rho(x,y))d\bdq^n(y)\\
        & \geq \int_{\|y\|\leq M} \exp(-B_0 \rho^*)d\bdq^n(y)\\
        &\geq \frac{1}{2} \exp(-B_0 \rho^*)\triangleq \delta_0.
    \end{aligned}
\end{equation*}
Here, $B_0$ is the upper bound of $\tilde{\beta}^{n}$, since it is convergent. And $M$ is larger than the constant $A$ in Theorem \ref{theorem1}. $\rho^*$ is a constant equals to $\max_{x,y\in[-M,M]^d}\rho(x,y)$.
Using the same derivation as \eqref{conv}, we have 
\begin{equation*}
     \int \exp(-\tilde{\beta}^{n}\rho(x,y))d\bdq^n(y)-\int \exp(-\beta^*\rho(x,y))d\bdq^n(y)\rightrightarrows 0.
\end{equation*}
Thus,
\begin{equation*}
    \int_{\|x\|\leq M}\frac{|\int \exp(-\beta^*\rho(x,y))d\bdq^n(y)-\int \exp(-\tilde{\beta}^{n}\rho(x,y))d\bdq^n(y)|}{\delta_0}dp(x)\leq \varepsilon,\text{ when n is sufficiently large}
\end{equation*}
Meanwhile, we denote $g(x)=\max_{\|y\|\leq A}\rho(x,y)$, then
\begin{equation*}
    \begin{aligned}
        &-\int_{\|x\|> M} \log \frac{\int \exp(-\tilde{\beta}^{n}\rho(x,y))d\bdq^n(y)}{\int \exp(-\beta^*\rho(x,y))d\bdq^n(y)}dp(x)\\
        &=-\int_{\|x\|> M} \log\bigg(\int \exp(-\tilde{\beta}^{n}\rho(x,y))d\bdq^n(y)\bigg)dp(x)-\int_{\|x\|> M} \log\bigg(\int \exp(-{\beta}^{*}\rho(x,y))d\bdq^n(y)\bigg)dp(x)\\
        &\leq\int_{\|x\|> M} \bigg[-\log\bigg(\int_{\|y\|\leq A} \exp(-\tilde{\beta}^{n}\rho(x,y))d\bdq^n(y)\bigg)- \log\bigg(\int_{\|y\|\leq A} \exp(-{\beta}^{*}\rho(x,y))d\bdq^n(y)\bigg)\bigg]dp(x)\\
        &\leq\int_{\|x\|> M} \bigg[-\log\bigg(\int_{\|y\|\leq A} \exp(-\tilde{\beta}^{n}g(x))d\bdq^n(y)\bigg)- \log\bigg(\int_{\|y\|\leq A} \exp(-{\beta}^{*}g(x))d\bdq^n(y)\bigg)\bigg]dp(x)\\
        &\leq \int_{\|x\|> M} \bigg[-\log\bigg(1/2\ \exp(-\tilde{\beta}^{n}g(x))\bigg)- \log\bigg(1/2\  \exp(-{\beta}^{*}g(x))\bigg)\bigg]dp(x)\\
        &=\int_{\|x\|> M} \bigg[2\log 2+\tilde{\beta}^{n}g(x)+\beta^* g(x)\bigg]dp(x)\\
        &\leq \int_{\|x\|> M} \bigg[2\log 2+B_0 g(x)+\beta^* g(x)\bigg]dp(x)
    \end{aligned}
\end{equation*}
Combining the two parts and taking the limit $n\rightarrow\infty$, we obtain
\begin{equation*}
    \begin{aligned}
        &\limsup_n [F(\bdq^n,\tilde{\beta}^{n})-F(\bdq^n,\beta^*)]=\limsup_n -\int \log \frac{\int \exp(-\tilde{\beta}^{n}\rho(x,y))d\bdq^n(y)}{\int \exp(-\beta^*\rho(x,y))d\bdq^n(y)}dp(x)\\
        &=\limsup_n \bigg(-\int_{\|y\|\leq M} \log \frac{\int \exp(-\tilde{\beta}^{n}\rho(x,y))d\bdq^n(y)}{\int \exp(-\beta^*\rho(x,y))d\bdq^n(y)}dp(x)-\int_{\|y\|>M} \log \frac{\int \exp(-\tilde{\beta}^{n}\rho(x,y))d\bdq^n(y)}{\int \exp(-\beta^*\rho(x,y))d\bdq^n(y)}dp(x)\bigg)\\
        &\leq 0+\int_{\|x\|> M} \bigg[2\log 2+B_0 g(x)+\beta^* g(x)\bigg]dp(x)
    \end{aligned}
\end{equation*}
Then by taking the limit of $M\rightarrow\infty$, we obtain the convergence. Here we used Assumption \eqref{assump_b} $\int g(x)dp(x)<\infty$.

 To prove the convergence of the solutions, we let $(\tilde{\bm{r}},\tilde{\beta})$ be a accumulation point of the solution sequence 
    $\{(\bm{r}^n,\beta^n)\}_{n=1}^{\infty}$. 
    Then there is a subsequence
    $\{(\bm{r}^{n_k},\beta^{n_k})\}_{k=1}^{\infty}$ satisfying 
    $\bm{r}^{n_k}\rightarrow \tilde{\bm{r}}$ and $\beta^{n_k}\rightarrow \tilde{\beta}$. Next, we have 
    $$F(\tilde{\bm{r}},\tilde{\beta})=\lim_k F(\bm{r}^{n_k},\beta^{n_k})=F(\bm{r}^*,\beta^*),$$
    since we have proven $\lim_n F(\bm{r}^{n},\beta^n)=F(\bm{r}^*,\beta^*)$.
    By the optimal property of $(\bm{r}^{*},\beta^{*}),(\bm{r}^{n_k},\beta^{n_k})$, we have
    \begin{equation*}
F(\tilde{\bm{r}},\tilde{\beta})=F(\bm{r}^{*},\beta^{*})\leq F(r^{n_k},\beta^{^*}) \leq F(\bm{r}^{n_k},\beta^{n_k})\leq F(\tilde{q}^{n_k},\beta^{n_k}).
    \end{equation*}
    Here, $\tilde{q}^{n_k}\in W_{n_k}$ is a discrete version of $\Bar{r}=\text{argmin}_{r\in W} F(r,\tilde{\beta})$.
    Similar to the value convergence analysis above, we have $F(\tilde{q}^{n_k},\beta^{n_k})\rightarrow F(\Bar{r},\tilde{\beta})$.
    Thus, we have
    \begin{equation*}
        F(\tilde{\bm{r}},\tilde{\beta})\leq F(\Bar{r},\tilde{\beta})\leq F(r,\tilde{\beta}),\ \forall r\in W.
    \end{equation*}
    This means $\tilde{\bm{r}}=\text{argmin}_{r\in W} F(r,\tilde{\beta})$ and
    \begin{equation*}
F(\tilde{\bm{r}},\tilde{\beta})=\min_{r\in W} F(r,\tilde{\beta})\triangleq h(\tilde{\beta}).
    \end{equation*}
    However, $F(\tilde{\bm{r}},\tilde{\beta})=F(\bm{r}^{*},\beta^{*})=h(\beta^*)$, and $h(\beta^*)=\max_{\beta\geq 0} h(\beta)$, due to the optimal property of $(\bm{r}^{*},\beta^{*})$. So, we obtain $\tilde{\beta}$ is optimal, \textit{i.e.}, $\tilde{\beta}=\text{argmax}_{\beta\geq 0}\ h(\beta)$.
And combining $\tilde{\bm{r}}=\text{argmin}_{r\in W} F(r,\tilde{\beta})$, we obtain $(\tilde{\bm{r}},\tilde{\beta})$ is optimal.
     \end{proofs}

\section{Convergence Rate and Algorithm Complexity}

First, we will establish some convergence rate results of problem \eqref{RDd}. Before the analysis of the convergence rate, we need a lemma first.
\begin{lemma}\label{Lip1}
    $e^{-\lambda \rho(x,y)}$ is Lipschitz continuous, when $(x,y)\in [-M,M]^d\times [-M,M]^d$, \textit{i.e.}, there exists a constant $L>0$ satisfying:
    \begin{equation*}
        |e^{-\lambda \rho(x_1,y_1)}-e^{-\lambda \rho(x_2,y_2)}|\leq L(\|x_1-x_2\|+\|y_1-y_2\|),\forall (x_1,y_1),(x_2,y_2)\in [-M,M]^d\times [-M,M]^d.
    \end{equation*}
\end{lemma}
\begin{proofs}
    Since $e^{-\lambda \rho(x,y)}$ is continuously differentiable, its Jacobi matrix $J(x,y)$ is bounded in $[-M,M]^d\times [-M,M]^d$, \textit{i.e.}, $\|J(x,y)\|\leq L$. Then by the Finite Increment Theorem, we have
    \begin{equation*}
        |e^{-\lambda \rho(x_1,y_1)}-e^{-\lambda \rho(x_2,y_2)}|\leq \|J(x,y)\|(\|x_1-x_2\|+\|y_1-y_2\|),
    \end{equation*}
    for some $(x,y)\in [-M,M]^d\times [-M,M]^d$. Then by the bound on $\|J(x,y)\|$, we obtain the Lipchitz continuous property.
\end{proofs}

Next, we will prove the convergence rate of the discrete schemes \eqref{RDd}.
\begin{theorems}\label{theorem12}
    When the distribution $p$ of $X$ is supported in $[-M,M]^d$, 
    the optimal values $f(\bm{r}^n)$ of the discrete problem \eqref{RDd} satisfy the error estimate
    \begin{equation*}
        |f(\bm{r}^n)-f^*|\leq Ch,
    \end{equation*}
    where $C$ is a constant and $h=2M/n^{\frac{1}{d}}$ is the discretization step size.
\end{theorems}
\begin{proofs}
By Lemma 4.1 in \cite{fix1978rate}, we have the distribution $\bdr$ of the reproduction variable is supported in $[-M,M]^d$ as long as $\rho(x,y)$ is a strictly increasing continuous difference distortion measure.
Thus, we can just take the equidistant discretization nodes $\{y_j^n\}_{j=1}^n$ from $[-M,M]^d$.
    Due to the inequality \eqref{f} in Theorem \ref{theorem1}, we have 
    \begin{equation*}
        0\leq f(\bdr^n)-f(\bdr^*)\leq f(\bdq^n)-f(\bdr^*). 
    \end{equation*}
    Thus, we only need to evaluate $f(\bdq^n)-f(\bdr^*)$.
    \begin{equation*}
        \begin{aligned}
            &\bigg|\int \exp(-\lambda \rho(x,y))d\bdq^n(y)- \int \exp(-\lambda \rho(x,y))d\bdr^*(y)\bigg|\\
            &=\bigg|\sum_i\int_{I_i} d\bdr^*(y)\exp(-\lambda \rho(x,y_i^n))- \int \exp(-\lambda \rho(x,y))d\bdr^*(y)\bigg|\\
            &=\bigg|\sum_i\int_{I_i} d\bdr^*(y)\exp(-\lambda \rho(x,y_i^n))- \sum_i\int_{I_i} \exp(-\lambda \rho(x,y))d\bdr^*(y)\bigg|\\
            &\leq \sum_i\int_{I_i} |\exp(-\lambda \rho(x,y))-\exp(-\lambda \rho(x,y_i^n))|d\bdr^*(y)\\
            &\leq \sum_i\int_{I_i} L\|y-y_i^n\|d\bdr^*(y)\\
            &\leq \sum_i\int_{I_i} Lh/2\ d\bdr^*(y)=Lh/2\\
        \end{aligned}
    \end{equation*}
    The second equality is due to $\bigcup_{i=1}^n I_i\supseteq [-M,M]^d$ and we have used the result in Lemma \ref{Lip1}.
    
    Next, we have an evaluation on the lower bound of $\int \exp(-\lambda \rho(x,y))d\bdq^n(y)$.
    \begin{equation*}
    \begin{aligned}
           \int \exp(-\lambda \rho(x,y))d\bdq^n(y)&=\int_{\|y\|\leq M} \exp(-\lambda \rho(x,y))d\bdq^n(y)\\
        &\geq e^{-\lambda \rho^*}\int_{\|y\|\leq M} d\bdq^n(y)=e^{-\lambda \rho^*}\triangleq \delta_0,\forall x\in [-M,M]^d.
    \end{aligned}
    \end{equation*}
    Here, $\rho^*=\max_{x,y\in [-M,M]^d} \rho(x,y)$.
    Similarly, we have
    \begin{equation*}
        \int \exp(-\lambda \rho(x,y))d\bdr^*(y)\geq \delta_0.
    \end{equation*}

    Then,
     \begin{equation*}
        \begin{aligned}
            &\log\bigg(\int \exp(-\lambda \rho(x,y))d\bdq^n(y)\bigg)- \log\bigg(\int \exp(-\lambda \rho(x,y))d\bdr^*(y)\bigg)\\
            &=\log\bigg(\big(\int \exp(-\lambda \rho(x,y))d\bdq^n(y)-\int \exp(-\lambda \rho(x,y))d\bdr^*(y)\big)\bigg/\int \exp(-\lambda \rho(x,y))d\bdr^*(y)+1\bigg)\\
            &\leq \big(\int \exp(-\lambda \rho(x,y))d\bdq^n(y)-\int \exp(-\lambda \rho(x,y))d\bdr^*(y)\big)\bigg/\int \exp(-\lambda \rho(x,y))d\bdr^*(y)\\
            &\leq Lh/(2\delta_0)\\
        \end{aligned}
    \end{equation*}

    Finally, we have an evaluation on the convergence rate
    \begin{equation*}
        \begin{aligned}
            f(\bdq^n)-f(\bdr^*)&=\int \log\bigg(\int \exp(-\lambda \rho(x,y))d\bdq^n(y)\bigg)dp(x)-\int \log\bigg(\int \exp(-\lambda \rho(x,y))d\bdr^*(y)\bigg)dp(x)\\
            &=\int_{\|x\|\leq M}\bigg(\log\bigg(\int \exp(-\lambda \rho(x,y))d\bdq^n(y)\bigg)- \log\bigg(\int \exp(-\lambda \rho(x,y))d\bdr^*(y)\bigg)\bigg)dp(x)\\
            &\leq \int_{\|x\|\leq M} Lh/(2\delta_0)\  dp(x)\\
            &\leq Lh/(2\delta_0)= O(h).
        \end{aligned}
    \end{equation*}
    Thus, 
    \begin{equation*}
        f(\bdr^n)-f(\bdr^*)=O(h).
    \end{equation*}
    Since the total number of discrete nodes is $n$, there is $n^{\frac{1}{d}}$ nodes in each direction. So, the discretization distant is $\frac{2M}{n^{1/d}}$. Thus the convergence rate equals $O(1/n^{\frac{1}{d}})$ as well.
\end{proofs}

Next, based on the convergence rate analysis above, we can conduct a complexity analysis of solving the continuous RD problem \eqref{RD} for achieving $\varepsilon$-accuracy via the BA algorithm.
\begin{theorems}
    To ensure $\varepsilon$-accuracy when computing the optimal value, the BA algorithm needs $O(\frac{m|\log\varepsilon|}{\varepsilon^{d+1}})$ arithmetic operations. Here, $m$ is the number of discretizaion nodes of $X$ when conducting numerical integration and $d$ is the dimension of $\mathcal{Y}$.
\end{theorems}
\begin{proofs}
    By Theorem \ref{theorem12}, to ensure $f(\bdr^n)-f(\bdr^*)\leq \varepsilon$, we need $n$ to satisfy $1/n^{1/d}\sim \varepsilon$, \textit{i.e.}, $n\sim 1/\varepsilon^d$.
    Then we use the BA algorithm to solve the associated discrete problem within $\varepsilon$ tolerance.
    As shown in \cite[Theorem 2, Equation (96)]{hayashi2023bregman}, the BA algorithm needs $O(\log n/\varepsilon)$ iterations to achieve $\varepsilon$-accuracy. 
    In each iteration, the BA algorithm iterates between two variables $w(y_i^n|x)$ and $r(y_i^n)$ in the following way:
    \begin{equation*}
        \begin{aligned}
            w(y_i^n|x)&=\bigg(r(y_i^n)e^{-\beta \rho(x,y_i^n)}\bigg)\bigg/\sum_{i=1}^n e^{-\beta \rho(x,y_i^n)}r(y_i^n),\\
            r(y_i^n)&=\int p(x)w(y_i^n|x)dx.
        \end{aligned}
    \end{equation*}
    Let $x_1,x_2\cdots x_m$ be the nodes of numerical integration with respect to $x$. When computing $w(y_i^n|x)$, we need to perform matrix-vector multiplication $$\sum_{i=1}^n e^{-\beta \rho(x_j,y_i^n)}r(y_i^n),\ j=1,2\cdots m,$$
    and it involves $O(mn)$ arithmetic operations. The total computation cost for $w(y_i^n|x)$ is $O(mn)$.
    When computing $r(y_i^n)$, we need to perform numerical integration,
    \begin{equation*}
        \int p(x)w(y_i^n|x)dx\sim \sum_{j=1}^m A_j p(x_j)w(y_i^n|x_j),\ i=1,2\cdots n,
    \end{equation*}
    where $A_j$ are the numerical integration coefficients. This is a matrix-vector multiplication and it involves $O(mn)$ arithmetic operations. Thus, the computation cost for each iteration in the BA algorithm is $O(mn)$.
    Hence, the total computation cost is $O(mn\log n/\varepsilon)=O(m|\log\varepsilon|/\varepsilon^{d+1})$.
\end{proofs}

Correspondingly, the convergence rate of problem \eqref{RD1d} can be obtained.

\begin{theorems}\label{theorem21}
    When the distribution $p$ of $X$ is supported in $[-M,M]^d$, 
    the optimal values $F(\bm{r}^n,\beta^n)$ of discrete problem \eqref{RD1d} satisfy the error estimate
    \begin{equation*}
        |F(\bm{r}^n,\beta^n)-F^*|\leq Ch,
    \end{equation*}
    where $C$ is a constant and $h=2M/n^{\frac{1}{d}}$ is the discretization step size.
\end{theorems}

Before the analysis of the convergence rate, we need a lemma first.
\begin{lemma}
    $e^{-\beta \rho(x,y)},\ e^{-\beta \rho(x,y)}\rho(x,y)$ is uniformly Lipschitz continuous for $0<B_1\leq\beta\leq B_0$, when $(x,y)\in [-M,M]^d\times [-M,M]^d$, \textit{i.e.}, there exists a constant $L>0$ satisfying:
    \begin{equation*}
    \begin{aligned}
         |e^{-\beta \rho(x_1,y_1)}-e^{-\beta \rho(x_2,y_2)}|\leq L(\|x_1-x_2\|+\|y_1-y_2\|),\\
         |e^{-\beta \rho(x_1,y_1)}\rho(x_1,y_1)-e^{-\beta \rho(x_2,y_2)}\rho(x_2,y_2)|\leq L(\|x_1-x_2\|+\|y_1-y_2\|),\\
        \forall (x_1,y_1),(x_2,y_2)\in [-M,M]^d\times [-M,M]^d,\quad \forall 0<B_1\leq\beta\leq B_0,
    \end{aligned}
    \end{equation*}
\end{lemma}
\begin{proofs}
    Since $e^{-\beta \rho(x,y)}$ is continuously differentiable, its Jacobi matrix $J_{\beta}(x,y)$ with respect to $(x,y)$ is continuous in $[-M,M]^d\times [-M,M]^d$. And taking into account $\beta$, $J_{\beta}(x,y)$ is continuous in $[-M,M]^d\times [-M,M]^d\times [B_1,B_0]$. Thus it is bounded, \textit{i.e.}, $\|J_{\beta}(x,y)\|\leq L$. Then by the Finite Increment Theorem, we have
    \begin{equation*}
        |e^{-\beta \rho(x_1,y_1)}-e^{-\beta \rho(x_2,y_2)}|\leq \|J_{\beta}(x,y)\|(\|x_1-x_2\|+\|y_1-y_2\|),
    \end{equation*}
    for some $(x,y)\in [-M,M]^d\times [-M,M]^d$. Then by the bound on $\|J_{\beta}(x,y)\|$, we obtain the Lipschitz continuous property. Similarly, we can prove the Lipschitz continuous property of $e^{-\beta \rho(x,y)}\rho(x,y)$.
\end{proofs}

Next, we will give the proof of Theorem \ref{theorem21}.

\begin{proofs}
    Since $r$ is supported in $[-M,M]^d$, we can just take the equidistant discretization nodes $\{y_j^n\}_{j=1}^n$ from $[-M,M]^d$.
    Due to the inequality \eqref{f1} in Theorem 4, we have 
    \begin{equation*}
        0\leq f(\bm{r}^n,\beta^n)-f(\bm{r}^*,\beta^*)\leq f(\bm{q}^n,\tilde{\beta}^n)-f(\bm{r}^*,\beta^*). 
    \end{equation*}
    Thus, we only need to evaluate $f(\bm{q}^n,\tilde{\beta}^n)-f(\bm{r}^*,\beta^*)$.
    
    \textbf{First, we estimate the convergence rate of $\tilde{\beta}^n\rightarrow \beta^*$}.
    \begin{equation*}
        \begin{aligned}
            &\bigg|\int \exp(-\tilde{\beta}^n \rho(x,y))d\bdq^n(y)- \int \exp(-\tilde{\beta}^n \rho(x,y))d\bdr^*(y)\bigg|\\
            &=\bigg|\sum_i\int_{I_i} d\bdr^*(y)\exp(-\tilde{\beta}^n \rho(x,y_i^n))- \int \exp(-\tilde{\beta}^n \rho(x,y))d\bdr^*(y)\bigg|\\
            &=\bigg|\sum_i\int_{I_i} d\bdr^*(y)\exp(-\tilde{\beta}^n \rho(x,y_i^n))- \sum_i\int_{I_i} \exp(-\tilde{\beta}^n \rho(x,y))d\bdr^*(y)\bigg|\\
            &\leq \sum_i\int_{I_i} |\exp(-\tilde{\beta}^n \rho(x,y))-\exp(-\tilde{\beta}^n \rho(x,y_i^n))|d\bdr^*(y)\\
            &\leq \sum_i\int_{I_i} L\|y-y_i^n\|d\bdr^*(y)\\
            &\leq \sum_i\int_{I_i} Lh/2\ d\bdr^*(y)=Lh/2\\
        \end{aligned}
    \end{equation*}
    The second equality is due to $\bigcup_{i=1}^n I_i\supseteq [-M,M]^d$. And the second inequality uses the bound on $\tilde{\beta}^n$, \textit{i.e.}, $0<B_1\leq\tilde{\beta}^n\leq B_0$.
    Similarly, we have
    \begin{equation*}
        \bigg|\int \exp(-\tilde{\beta}^n \rho(x,y))\rho(x,y)d\bdq^n(y)- \int \exp(-\tilde{\beta}^n \rho(x,y))\rho(x,y)d\bdr^*(y)\bigg|\leq Lh/2.
    \end{equation*}
     Next, we have an evaluation on the lower bound of $\int \exp(-\tilde{\beta}^n \rho(x,y))d\bdq^n(y)$.
    \begin{equation}\label{low_delta}
    \begin{aligned}
           \int \exp(-\tilde{\beta}^n \rho(x,y))d\bdq^n(y)&=\int_{\|y\|\leq M} \exp(-\tilde{\beta}^n \rho(x,y))d\bdq^n(y)\\
        &\geq e^{-\tilde{\beta}^n \rho^*}\int_{\|y\|\leq M} d\bdq^n(y)=e^{-\tilde{\beta}^n \rho^*}\triangleq \delta_0,\forall x\in [-M,M]^d.
    \end{aligned}
    \end{equation}
    Here, $\rho^*=\max_{x,y\in [-M,M]^d} \rho(x,y)$.
    Similarly, we have
    \begin{equation*}
        \int \exp(-\tilde{\beta}^n \rho(x,y))d\bdr^*(y)\geq \delta_0.
    \end{equation*}
    And we have an upper bound estimation on $\int \exp(-\tilde{\beta}^n \rho(x,y))\rho(x,y)d\bdr^*(y)$.
    \begin{equation*}
        \begin{aligned}
            \int \exp(-\tilde{\beta}^n \rho(x,y))\rho(x,y)d\bdr^*(y)&\leq \int \exp(-B_1 \rho(x,y))\rho(x,y)d\bdr^*(y)\\
            &\leq \int M_1 d\bdr^*(y)=M_1.
        \end{aligned}
    \end{equation*}
    Here, $M_1$ is the upper bound on the function $\exp(-B_1 t)t,t\geq 0$.
    Now, since $G_{\bdr^*}(\beta^*)=0=G_{\bdq^n}(\tilde{\beta}^n)$, we have the following estimation
    \begin{equation*}
        \begin{aligned}
            &|G_{\bdr^*}(\tilde{\beta}^n)-G_{\bdr^*}(\beta^*)|=|G_{\bdr^*}(\tilde{\beta}^n)-G_{\bdq^n}(\tilde{\beta}^n)|\\
            &\leq  \int_{\|x\|\leq M}\bigg|\frac{\bigg(\int \exp({-\tilde{\beta}^n \rho(x,y)})\rho(x,y)d\bdq^n(y)\bigg)}{\bigg(\int \exp({-\tilde{\beta}^n \rho(x,y)})d\bdq^n(y)\bigg)}-\frac{\bigg(\int \exp({-\tilde{\beta}^n \rho(x,y)})\rho(x,y)d\bdr^*(y)\bigg)}{\bigg(\int \exp({-\tilde{\beta}^n \rho(x,y)})d\bdr^*(y)\bigg)}\bigg|dp(x)\\
            &=\int_{\|x\|\leq M}\bigg|\frac{b_1}{a_1}-\frac{b_2}{a_2}\bigg|dp(x)=\int_{\|x\|\leq M}\frac{|a_2 b_1-a_1 b_2|}{a_1 a_2}dp(x)\\
            &\leq\int_{\|x\|\leq M}\frac{a_2|b_1-b_2|+b_2|a_2-a_1|}{a_1 a_2}dp(x)\leq \int_{\|x\|\leq M}Lh/(2\delta_0)+\frac{M_1 Lh/2}{\delta_0^2}dp(x)\\
            &\leq Lh/(2\delta_0)+\frac{M_1 Lh/2}{\delta_0^2}=O(h).
        \end{aligned}
    \end{equation*}
    Here, for simplicity, we use $a_1,a_2,b_1,b_2$ to represent the corresponding integration.
    
    Next, we will give the estimate of $\tilde{\beta}^n-\beta^*$.
    Let $-L_1=G_{\bdr^*}^{\prime}(\beta^*)<0$, then in a neighborhood $(-\delta_2+\beta^*,\delta_2+\beta^*)$ of $\beta^*$, we have $G_{\bdr^*}^{\prime}(\beta)\leq -L_1/2$ and $\tilde{\beta}^n$ is in the neighborhood when $n$ is sufficiently large. Therefore, by the Lagrange Mean Value Theorem, we have
    \begin{equation*}
        |G_{\bdr^*}(\tilde{\beta}^n)-G_{\bdr^*}(\beta^*)|=|G_{\bdr^*}^{\prime}(\zeta)|\ |\tilde{\beta}^n-\beta^*|\geq \frac{L_1}{2}|\tilde{\beta}^n-\beta^*|.
    \end{equation*}
Here, $\zeta$ is a real number between $\tilde{\beta}^n$ and $\beta^*$. Thus, we obtain
\begin{equation*}
    |\tilde{\beta}^n-\beta^*|\leq \frac{2}{L_1}|G_{\bdr^*}(\tilde{\beta}^n)-G_{\bdr^*}(\beta^*)|=O(h).
\end{equation*}

\textbf{Next, we will give the convergence rate of $f(\bm{q}^n,\tilde{\beta}^n)-f(\bm{r}^*,\beta^*)$.}
\begin{equation*}
    |f(\bm{q}^n,\tilde{\beta}^n)-f(\bm{r}^*,\beta^*)|\leq |f(\bm{q}^n,\tilde{\beta}^n)-f(\bdq^n,\beta^*)|+|f(\bm{q}^n,\beta^*)-f(\bm{r}^*,\beta^*)|.
\end{equation*}
For the first part, we have the following estimation
\begin{equation*}
    \begin{aligned}
        &0\leq f(\bm{q}^n,\tilde{\beta}^n)-f(\bdq^n,\beta^*)=-\int_{\|x\|\leq M} \log \frac{\int \exp(-\tilde{\beta}^{n}\rho(x,y))d\bdq^n(y)}{\int \exp(-\beta^*\rho(x,y))d\bdq^n(y)}dp(x)-(\tilde{\beta}^n-\beta^*)D\\
        &\leq\int_{\|x\|\leq M} \log \bigg(\frac{\int \exp(-\beta^*\rho(x,y))d\bdq^n(y)-\int \exp(-\tilde{\beta}^{n}\rho(x,y))d\bdq^n(y)}{\int \exp(-\tilde{\beta}^{n}\rho(x,y))d\bdq^n(y)}+1\bigg)dp(x)+O(h)\\
        &\leq \int_{\|x\|\leq M}\frac{|\int \exp(-\beta^*\rho(x,y))d\bdq^n(y)-\int \exp(-\tilde{\beta}^{n}\rho(x,y))d\bdq^n(y)|}{\int \exp(-\tilde{\beta}^{n}\rho(x,y))d\bdq^n(y)}dp(x)+O(h)\\
        &\leq \int_{\|x\|\leq M}\frac{|\int \exp(-\beta^*\rho(x,y))d\bdq^n(y)-\int \exp(-\tilde{\beta}^{n}\rho(x,y))d\bdq^n(y)|}{\delta_0}dp(x)+O(h)
    \end{aligned}
\end{equation*}
Here, $\delta_0$ is a lower bound as shown in \eqref{low_delta}.\\
\begin{equation*}
    \begin{aligned}
        &|\int \exp(-\beta^*\rho(x,y))d\bdq^n(y)-\int \exp(-\tilde{\beta}^{n}\rho(x,y))d\bdq^n(y)|\\  
        &\leq\int e^{-\beta^* \rho(x,y)}|e^{({\beta}^*-\tilde{\beta}^{n})\rho(x,y)}-1|d\bdq^n(y)\\
        &\leq\int_{\|y\|\leq M} |e^{({\beta}^*-\tilde{\beta}^{n})\rho(x,y)}-1|\ d\bdq^n(y)\\
        &\leq\int_{\|y\|\leq M} (e^{|{\beta}^*-\tilde{\beta}^{n}|\rho(x,y)}-1)\ d\bdq^n(y)\\
        &\leq\int_{\|y\|\leq M} (e^{|{\beta}^*-\tilde{\beta}^{n}|\rho^*}-1)\ d\bdq^n(y)\\
        &\leq (e^{|{\beta}^*-\tilde{\beta}^{n}|\rho^*}-1)\leq 2|{\beta}^*-\tilde{\beta}^{n}|\rho^*=O(h), \text{ when n is sufficiently large}
    \end{aligned}
\end{equation*}
Here, $\rho^*=\max_{x,y\in [-M,M]^d} \rho(x,y)$.
Thus,
\begin{equation*}
    \begin{aligned}
        |f(\bm{q}^n,\tilde{\beta}^n)-f(\bdq^n,\beta^*)|&\leq \int_{\|x\|\leq M}\frac{|\int \exp(-\beta^*\rho(x,y))d\bdq^n(y)-\int \exp(-\tilde{\beta}^{n}\rho(x,y))d\bdq^n(y)|}{\delta_0}dp(x)+O(h)\\
        &\leq  \int_{\|x\|\leq M}O(h)/{\delta_0}\ dp(x)+O(h)\leq O(h)/{\delta_0}+O(h)=O(h).
    \end{aligned}
\end{equation*}
Meanwhile, due to Theorem 2, $|f(\bdq^n,\beta^*)-f(\bdr^*,\beta^*)|=O(h)$. Thus, we obtain the convergence rate $|f(\bdr^n,\beta^n)-f(\bdr^*,\beta^*)|=O(h)=O(1/n^{\frac{1}{d}}).$
    
    \end{proofs}
    
Using the recently proposed CBA algorithm \cite{chen2023constrained}, we can solve the original RD problem \eqref{RD1} directly. The next theorem gives the complexity result of solving the continuous RD problem \eqref{RD1} for achieving $\varepsilon$-accuracy via the CBA algorithm.
\begin{theorems}
    To ensure $\varepsilon$-accuracy when computing the optimal value, the CBA algorithm needs $O\big(\frac{m|\log\varepsilon|}{\varepsilon^{d+1}}(1+\log|\log\varepsilon|)\big)$ arithmetic operations \footnote{The CBA algorithm takes an additional cost $\log|\log\varepsilon|$ to update the multiplier $\beta$, while the BA algorithm fixes $\beta$. Thus, the CBA algorithm is applicable for solving the original problem \eqref{RD1} directly.}. Here, $m$ is the number of discretizaion nodes of $X$ when conducting numerical integration and $d$ is the dimension of $\mathcal{Y}$.
\end{theorems}
\begin{proofs}
    By Theorem \ref{theorem21}, to ensure $F(\bdr^n,\beta^n)-F(\bdr^*,\beta^*)\leq \varepsilon$, we need $n$ to satisfy $1/n^{1/d}\sim \varepsilon$, \textit{i.e.}, $n\sim 1/\varepsilon^d$.
    Then we use the CBA algorithm to solve the discrete problem within $\varepsilon$ tolerance. As shown in a recent work \cite{chen2023constrained}, we need $O(\frac{mn\log n}{\varepsilon}(1+\log|\log\varepsilon|))$ arithmetic operations to achieve $\varepsilon$-accuracy. Thus, the complexity is
    \begin{equation*}
        O\big(\frac{mn\log n}{\varepsilon}(1+\log|\log\varepsilon|)\!\big)\!\!=\!O\big(\frac{m|\log\varepsilon|}{\varepsilon^{d+1}}(1+\log|\log\varepsilon|)\!\big).
    \end{equation*}
\end{proofs}

\section{Numerical Experiments}
In this section, we conduct experiments on uniform source to confirm the convergence. We consider the uniform source on interval $[-8,8]$ and conduct experiments with different discretization parameters, namely the node number $n=20, 40, 80, 160$ of $Y$, while we take the node number $m=300$ of $X$ to ensure computing the integral of $x$ with high accuracy. Moreover, we use the BA algorithm and the CBA algorithm to solve the discrete RD problem \eqref{RDd} and \eqref{RD1d} respectively with high accuracy. The corresponding results of the reproduction distribution are illustrated in Figure 1 and Figure 2.
\begin{figure}[ht]
    \centerline{\includegraphics[width=1\textwidth]{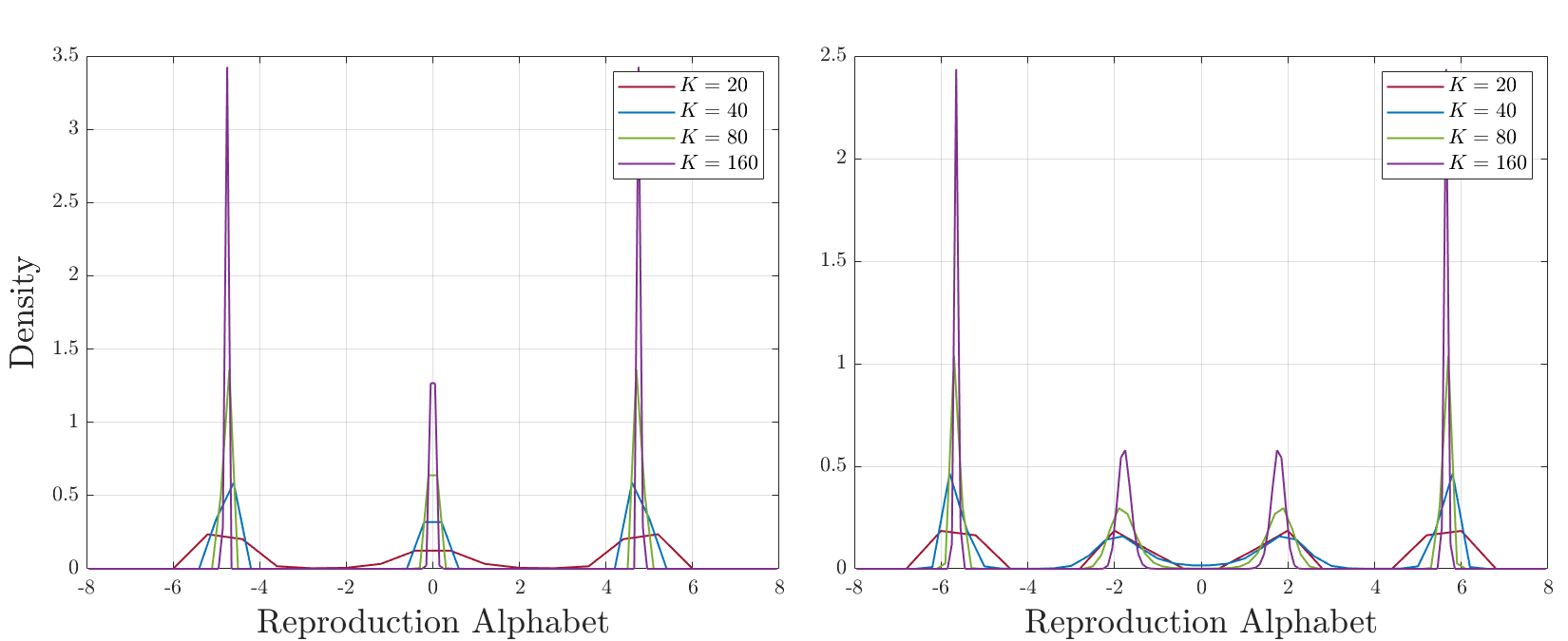}}
    \label{Fig: uniform1}
   \caption{The discrete optimal reproduction produced by the BA algorithm for the slope $\beta=0.1$ (left) and $\beta=0.2$ (right).}
\end{figure}
\begin{figure}[ht]
    \centerline{\includegraphics[width=1\textwidth]{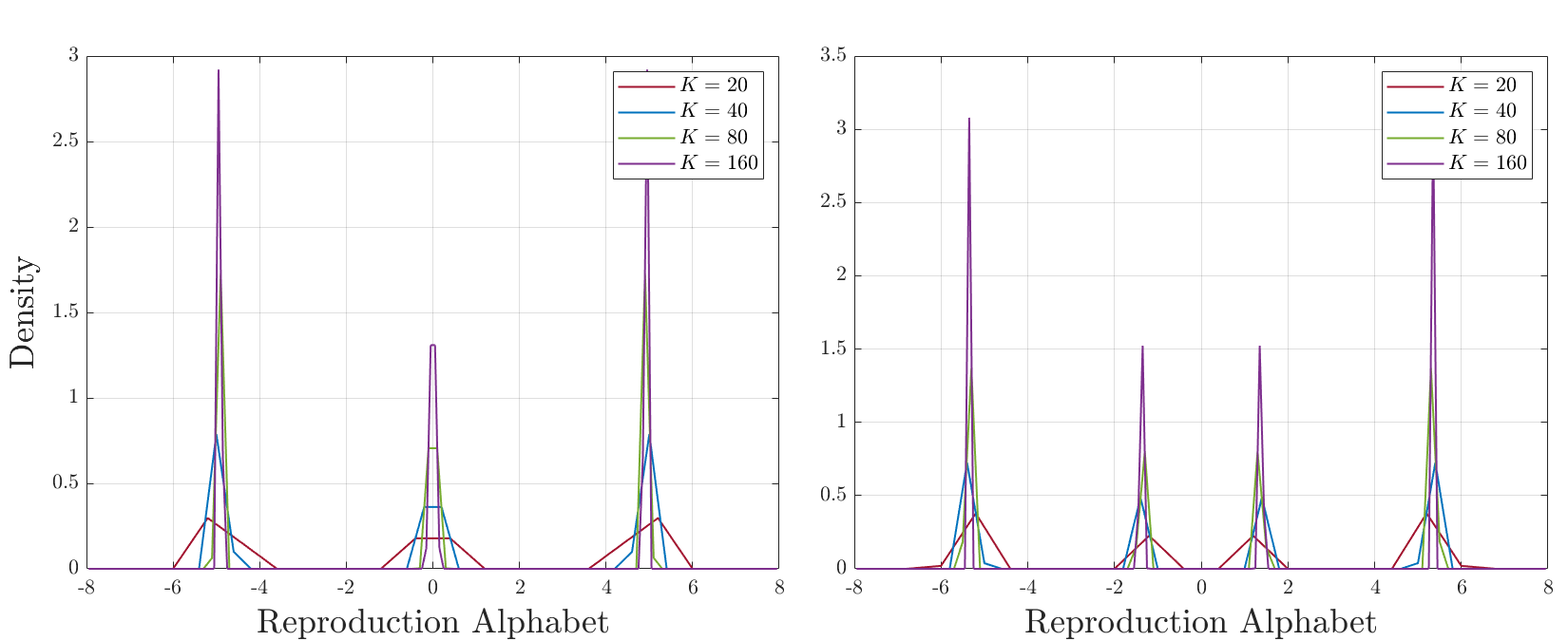}}
    \label{Fig: uniform2}
   \caption{The discrete optimal reproduction produced by the CBA algorithm for the distortion $D=4$ (left) and $D=3$ (right).}
\end{figure}

 As shown in \cite{rose94}, the optimal reproduction of the uniform source is a discrete distribution. 
From the figures, the convergence is clearly demonstrated and the solutions of discrete problems converge to a discrete distribution as the grids become finer.

\section{conclusion}  \label{sec_6}
In this paper, we prove the convergence of discrete schemes for computing the continuous RD problem and establish convergence rate results and complexity estimations.
Numerical experiments confirm the convergence. Considering the fundamental role of the RD problem, it is envisioned that our method may lead to a series of applications to various information problems.

\bibliographystyle{bibliography/IEEEtran}
\bibliography{bibliography/RD_REF}
\end{document}

%% file: notation.tex
\newcommand{\bdr}{\boldsymbol{r}}

\newcommand{\bdq}{\boldsymbol{q}}